\documentclass[letterpaper]{jpconf}

\usepackage[pdftex]{graphicx}
\usepackage{hyperref}
\usepackage{amsmath,amssymb,amsfonts} 
\usepackage{subfigure} 

\begin{document}

\DeclareGraphicsExtensions{.pdf, .jpg}

\title{Inclusive $A_{LL}$ Measurements at STAR}


\author{Adam Kocoloski (for the STAR Collaboration)}
\address{Massachusetts Institute of Technology\linebreak
  77 Massachusetts Ave., Cambridge, MA  02139}


\begin{abstract}
  One of the primary goals of the RHIC Spin program is to determine the gluon
  polarization distribution within the proton. At leading order, $pp$ collisions
  involve a mixture of quark-quark, quark-gluon, and gluon-gluon scattering. In
  RHIC, the gluon-gluon and quark-gluon contributions dominate, making the
  accelerator an ideal tool to explore gluon polarization. The STAR experiment
  has measured the longitudinal double-spin asymmetry $A_{LL}$ for inclusive
  production of jets and pions, and for charged pion production opposite a jet,
  at $\sqrt{s} = 200 GeV$. The results provide valuable new constraints on the
  gluon polarization in the proton when included in a next-to-leading-order
  global analysis. The current status of the STAR measurements and the plans for
  future measurements will be discussed.
\end{abstract}

\section{Introduction}

Collisions of polarized protons at the Relativistic Heavy Ion Collider (RHIC)
offer a window into the spin composition of the proton through measurements of
longitudinal double-spin asymmetries in a variety of final states
\cite{Bunce:2000uv}. Jet and pion production at STAR \cite{Ackermann:2002ad} is
dominated by gg and qg scattering, providing sensitivity to the polarized gluon
distribution $\Delta g(x,Q^{2})$ over a restricted kinematic region of $0.03 < x
< 0.3$.

One can write the longitudinal double-spin asymmetry $A_{LL}$ in terms of
experimental quantities as \begin{equation}A_{LL} = \frac{1}{P_{1}P_{2}}
\frac{N^{++} - RN^{+-}}{N^{++} + RN^{+-}},\end{equation} where $P_{1,2}$ are the
polarizations of the colliding proton beams, $N^{++}$ and $N^{+-}$ are the
identified particle yields when the proton helicities are aligned and
anti-aligned, and $R$ is the ratio of the beam luminosities in the two helicity
configurations. The beam polarizations at RHIC are measured every few hours
using a high-statistics Coulomb Nuclear Interference (CNI) polarimeter
\cite{Jinnouchi:2004up}, and the analyzing power of the CNI polarimeter is
normalized using a gas jet polarimeter \cite{Okada:2006dd}. The spin-dependent
relative luminosities are measured at STAR using the Beam Beam Counters (BBCs),
segmented scintillator annuli located up and downstream of the STAR interaction
region that provide full azimuthal coverage. Coincident signals in the two BBCs
define STAR's $pp$ minimum-bias trigger condition; this approach samples 87\% of
the non-singly diffractive scattering cross section \cite{Adams:2003kv}. The
BBCs can also be used to measure residual transverse beam polarization, which
manifests as an azimuthal asymmetry in the scintillator tile counts
\cite{Kiryluk:2005gg}. The minimum-bias trigger condition is necessary but not
sufficient to select events for the analyses reviewed here. Also required is an
transverse energy deposit in one of the electromagnetic calorimeters which
enhances STAR's sampling of hard scattering events.

\section{Jets}

Jets are reconstructed at STAR using a midpoint cone algorithm \cite{Blazey:2000qt} that clusters charged track and electromagnetic energy deposits within a cone of radius $R = \sqrt{\Delta \eta^2 + \Delta \phi^2} = 0.7$.  Earlier measurements used a smaller 0.4 cone radius because the Barrel Electromagnetic Calorimeter (BEMC) was not yet fully commissioned. Charged track momenta are measured by the Time Projection Chamber (TPC). Neutral energy deposits are measured by the Barrel and Endcap EMCs. Events are selected using a Jet Patch (JP) trigger algorithm that looks for a transverse energy deposition in fixed $\Delta \eta \times \Delta \phi = 1.0 \times 1.0$ patches of the BEMC above some threshold.  In the 2006 RHIC run, that threshold was primarily set at 8.3 GeV.

\begin{figure}
  \subfigure[inclusive jet $A_{LL}$]{
    \includegraphics[width=0.62\textwidth]{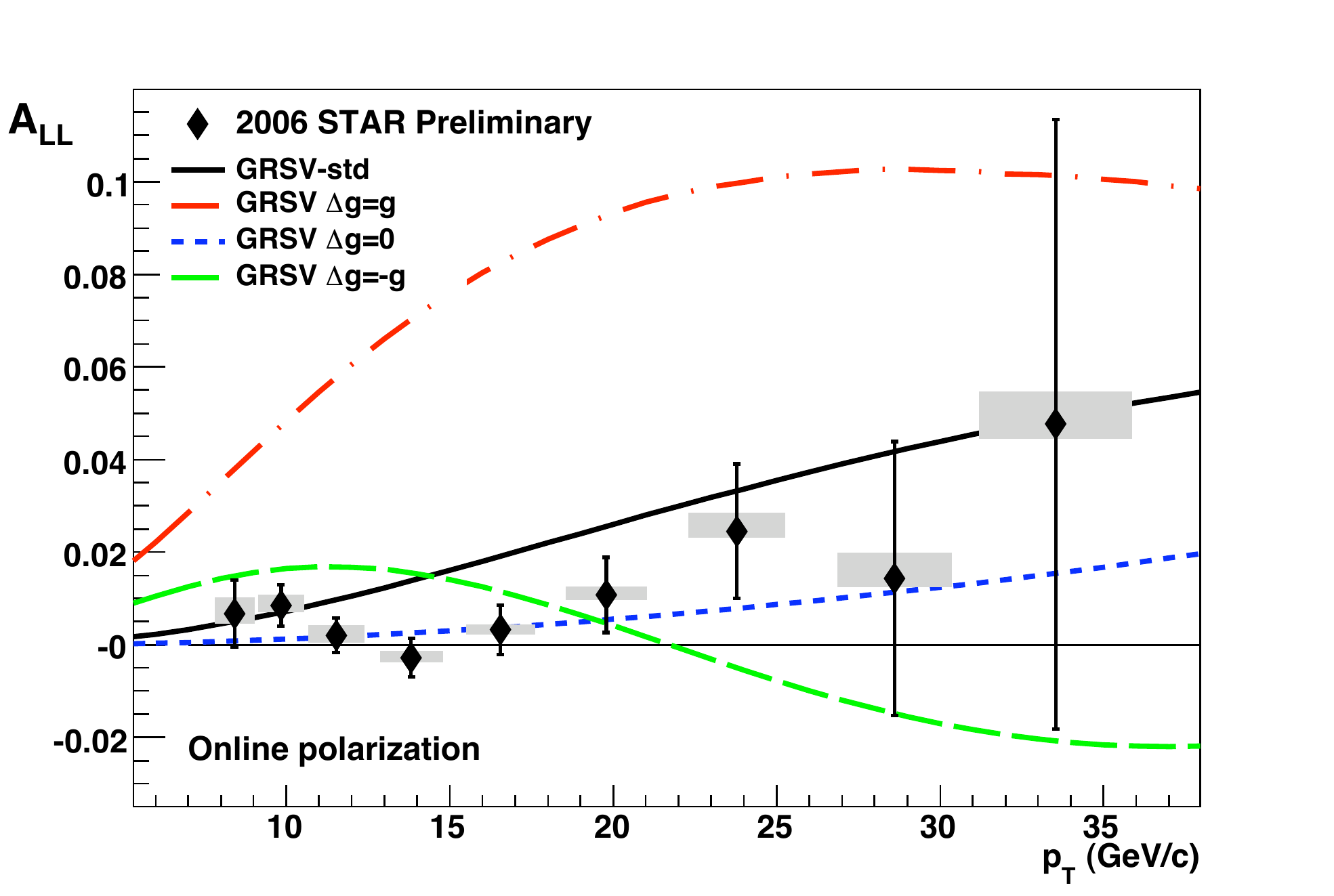}
    \label{fig:jet-all}
  }
  \subfigure[recent $\Delta g$ analysis]{
    \includegraphics[width=0.39\textwidth]{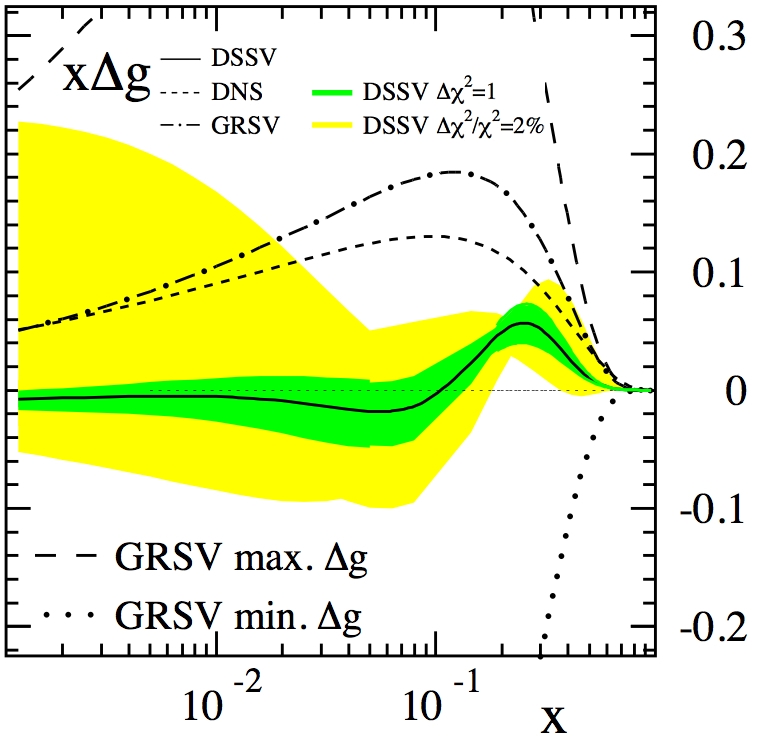}
    \label{fig:dssv}
  }
  \caption{Full jet reconstruction capabilities enable the flagship inclusive $A_{LL}$ measurement at STAR.  The left panel shows the most precise measurement of inclusive jet $A_{LL}$ to date, obtained using Run 6 data.  The right panel shows a recent global analysis of the polarized gluon polarization, the first such analysis to incorporate measurements from RHIC.}
  \label{fig:jet}
\end{figure}

Figure \ref{fig:jet-all} presents STAR's preliminary measurement of inclusive jet $A_{LL}$ obtained by analyzing 4.7 $pb^{-1}$ of data collected during the 2006 RHIC run \cite{Gagliardi:2008qw}.  The uncertainties on the data points are statistical; the gray bands plot asymmetric point-to-point systematic uncertainties.  The data are compared to a variety of predictions for $A_{LL}$ assuming various input distributions for the gluon polarization in the GRSV framework \cite{Jager:2004jh}.

The inclusive jet data from STAR (as well as $\pi^0$ data from PHENIX) have been incorporated into a global analysis used to extract polarized parton distribution functions \cite{deFlorian:2008mr}.  The right panel of Figure \ref{fig:jet} shows the result of this analysis for $\Delta g(x)$; compared to previous global analyses, the uncertainties on the pdf shrink dramatically  in the kinematic regime sampled by inclusive RHIC channels ($\sim0.03 < x < \sim0.3$).

\section{Neutral Pions}

Pions decaying via the two photon channel are reconstructed over a wide range in rapidity at STAR by a suite of electromagnetic calorimeters.  The Barrel and Endcap EMCs are lead-scintillator sampling calorimeters that use shower maximum detectors situated a few radiation lengths into the detector to resolve decays where both photons are coincident on the same tower, thus extending the $p_T$ reach of the measurements as high as 16 GeV/c.  The Forward Pion Detector (FPD) is comprised of lead-glass cell matrices situated $\sim$750 cm from the interaction region.  The FPD modules can be positioned at different transverse distances from the beamline, allowing for measurements at multiple values of $\eta$.

\begin{figure}
  \begin{center}
    \subfigure[BEMC]{
      \includegraphics[width=0.45\textwidth]{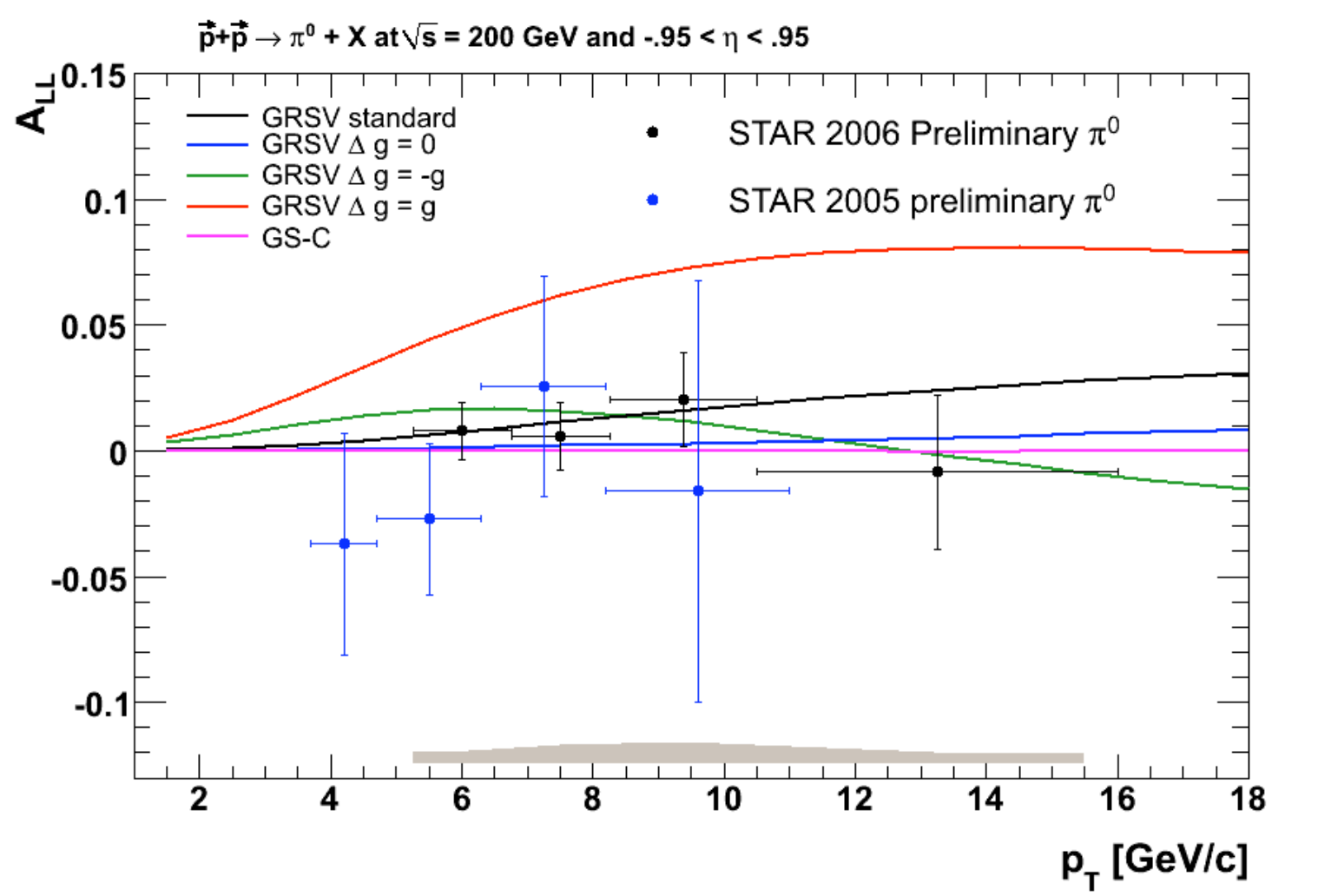}
      \label{fig:bemc-pi0}
    }
    \subfigure[EEMC]{
      \includegraphics[width=0.4\textwidth]{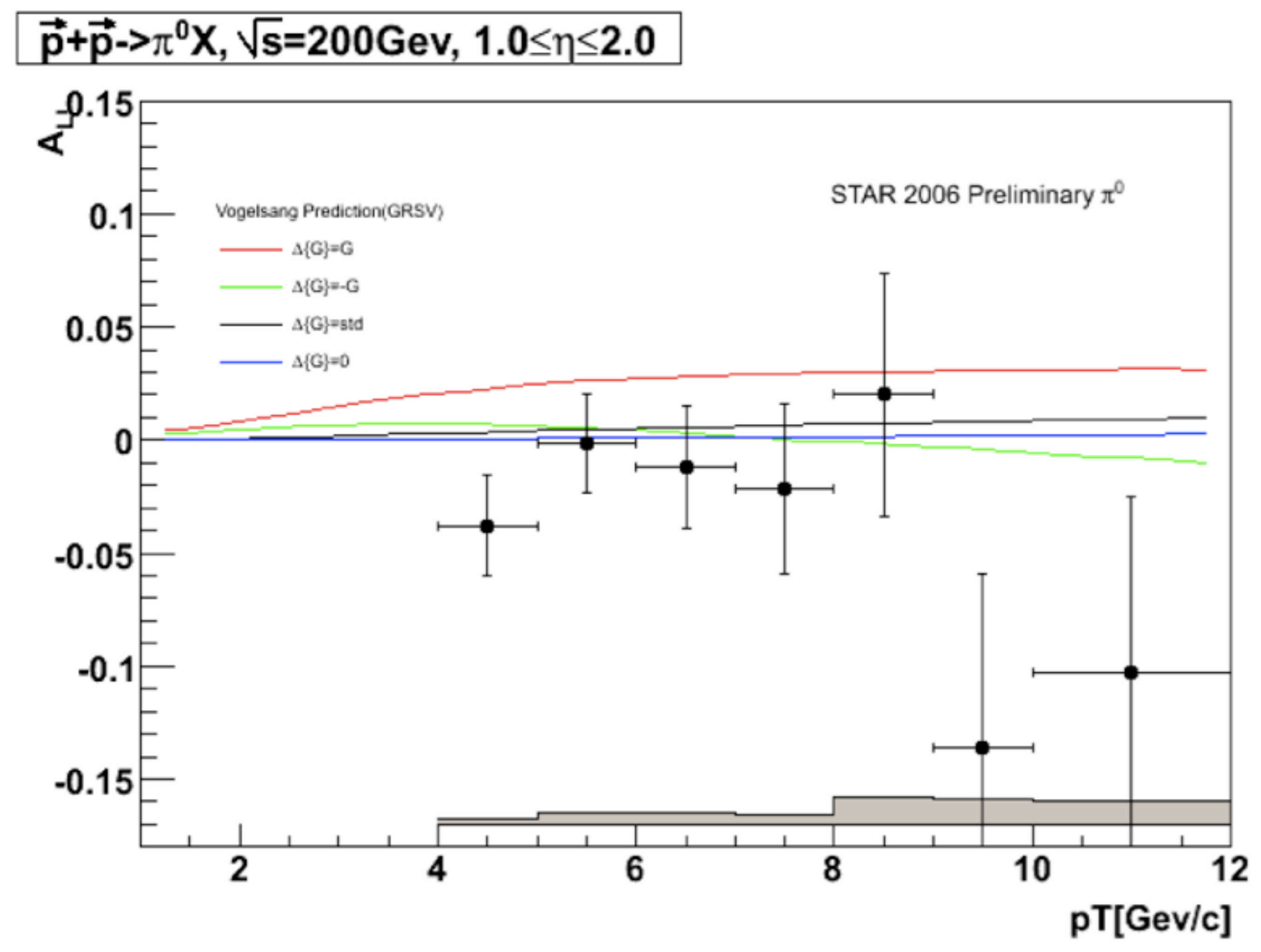}
      \label{fig:eemc-pi0}
    }
    \subfigure[FPD]{
      \includegraphics[width=0.37\textwidth]{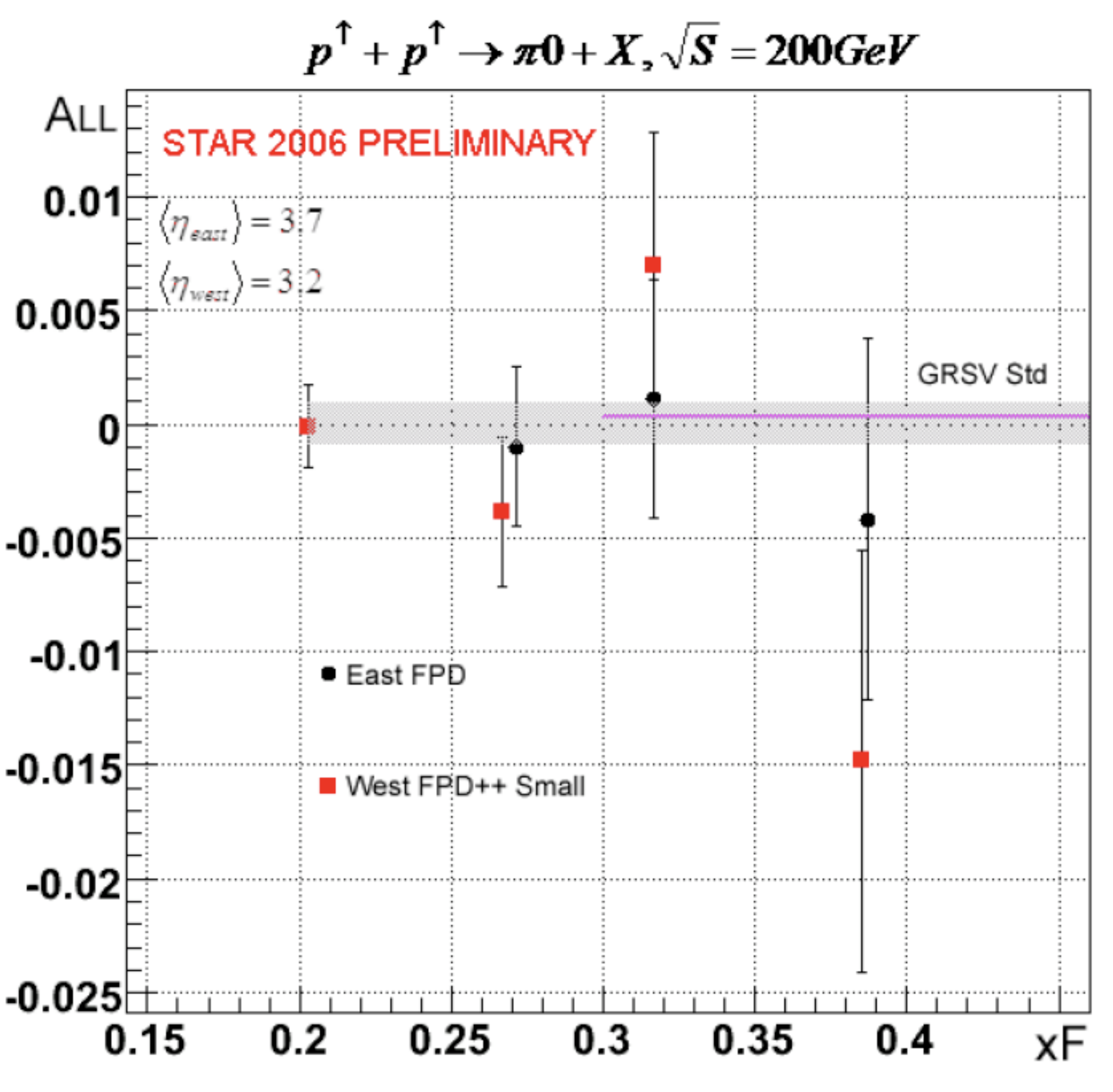}
      \label{fig:fpd-pi0}
    }
  \end{center}
  \caption{Asymmetries of inclusive neutral pion production extend over a wide range in rapidity at STAR and have a rather different systematic uncertainty profile than the inclusive jet $A_{LL}$, making them an excellent complementary set of measurements.}
  \label{fig:pi0-all}
\end{figure}

Figure \ref{fig:pi0-all} displays $A_{LL}$ results for each of these calorimeters \cite{Hoffman:2008zz, Wissink:2008zz}.  As in the case of the inclusive jet measurement, the data are compared to predictions assuming various input distributions for the polarized gluon distribution.  In contrast to the jet measurement, the bias from the triggering system in $\pi^0$ analyses is typically negligible.  A thorough understanding of $\pi^0$ reconstruction is also an important prerequisite for future direct photon measurements.


\section{Charged Pions}

\begin{figure}
  \includegraphics[width=\textwidth]{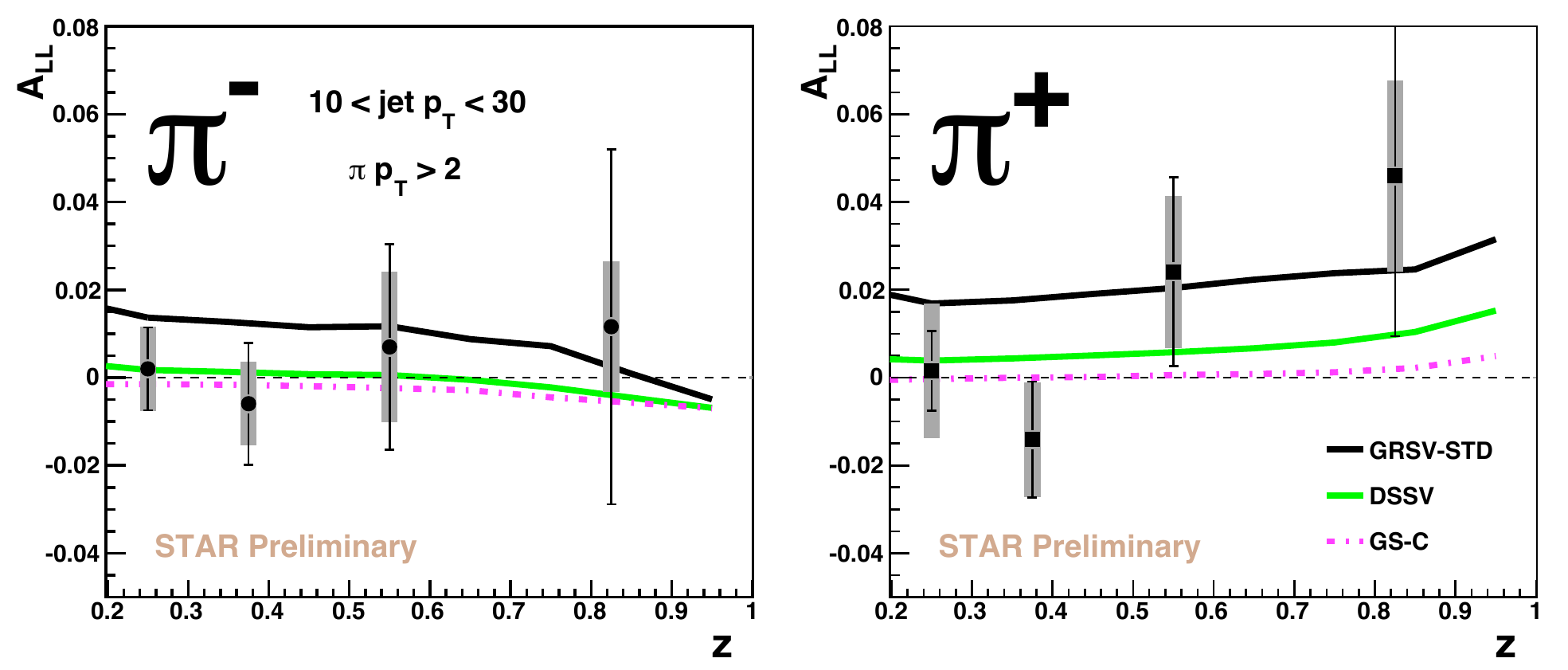}
  \caption{STAR's TPC allows for efficient reconstruction of charged pions over a wide range in transverse momentum, but the lack of a dedicated triggering system for charged pions makes an unbiased inclusive $A_{LL}$ a challenge.  The asymmetries shown here are obtained by triggering on a jet and counting pions on the opposite side of the event.}
  \label{fig:charged-all}
\end{figure}

Figure \ref{fig:charged-all} presents STAR's preliminary results for charged
pion $A_{LL}$ opposite a trigger jet, plotted versus $z \equiv
p_{T}(\pi)/p_{T}(jet)$ \cite{Kocoloski:2009aa}. Point-to-point systematic
uncertainties from a variety of sources are summed in quadrature and included as
the gray bars. The dominant systematic uncertainty arises from the use of the JP
trigger to select events. This trigger a) hardens the jet $p_{T}$ spectrum in
each $z$ bin, and b) preferentially selects quark jets over gluon jets. The
first effect is quantified by using Monte Carlo to determine the trigger
efficiency for jets at a given $p_{T}$, and the theoretical predictions for
$A_{LL}$ for this observable are adjusted to account for that efficiency
\cite{deFlorian:2009fw}. Future analyses with greater statistical precision may
be able to avoid this requirement by binning in both jet $p_{T}$ and $z$. For b)
STAR investigates how our LO Monte Carlo evaluation of $A_{LL}$ changes when we
require the JP trigger condition. Specifically, we compare our MC asymmetries
for the JP trigger with MC asymmetries that incorporate only the trigger
\textit{efficiency}, not the subprocess bias. The difference between the two is
a measure of how the trigger's preference for quark jets affects $A_{LL}$.

\section{Conclusions and Outlook}

STAR has a robust program of inclusive double-spin asymmetry measurements.  Global analyses incorporating STAR data have established significant new constraints on the polarized gluon distribution in the proton.  These measurements set the stage for a program of correlation measurements at both $\sqrt{s}$ = 200 GeV and 500 Gev that will extend the sampled $x$ range and more precisely map out the $x$-dependence of the polarized gluon distribution.

\section*{References}

\bibliographystyle{iopart-num}
\bibliography{kocolosk}
\end{document}